\documentclass[conference, letterpaper]{IEEEtran}
\usepackage[noadjust]{cite}

\usepackage{float}
\usepackage{notoccite}
\usepackage{dirtytalk}
\usepackage{csquotes}
\usepackage{url}
\usepackage{multirow,tabu}
\usepackage{hhline}
\usepackage{colortbl}
\usepackage[table,xcdraw]{xcolor}
\usepackage{adjustbox}
\usepackage[utf8x]{inputenc}
\usepackage[normalem]{ulem} 
\usepackage[colorinlistoftodos]{todonotes}
\usepackage{booktabs,threeparttable,stackengine}
\usepackage{longtable}
\usepackage{makecell}
\usepackage{array}
\newcolumntype{?}{!{\vrule width 1pt}} 

\makeatletter
\def\Cline#1#2{\@Cline#1#2\@nil}
\def\@Cline#1-#2#3\@nil{%
  \omit
  \@multicnt#1%
  \advance\@multispan\m@ne
  \ifnum\@multicnt=\@ne\@firhttps://www.sharelatex.com/project/59ce5a4336e6671a141981d4/output/output.pdf?compileGroup=standard&clsiserverid=clsi2-30&popupDownload=truestofone{&\omit}\fi
  \@multicnt#2%
  \advance\@multicnt-#1%
  \advance\@multispan\@ne
  \leaders\hrule\@height#3\hfill
  \cr}
\makeatother

\ifCLASSINFOpdf
\else
\fi
\ifCLASSINFOpdf
   \usepackage{graphicx} 
\else
\fi

\usepackage[cmex10]{amsmath}
\usepackage{color}
\usepackage{fancyhdr}
\usepackage[caption=false,font=footnotesize]{subfig}

\renewcommand{\thispagestyle}[2]{}

\fancypagestyle{plain}{
        \fancyhead{}
        \fancyhead[C]{first page center header}
        \fancyfoot{}
        \fancyfoot[C]{first page center footer}
}
\begin{document}

%
\title{The Impact of Quantum Computing on Present Cryptography}

\author{\IEEEauthorblockN{Vasileios Mavroeidis, Kamer Vishi, Mateusz D. Zych, Audun J\o sang}
\IEEEauthorblockA{Department of Informatics, University of Oslo, Norway\\
Email(s): \{vasileim, kamerv, mateusdz, josang\}@ifi.uio.no}}


%


\maketitle

\begin{abstract}
The aim of this paper is to elucidate the implications of quantum computing in present cryptography and to introduce the reader to basic post-quantum algorithms.
In particular the reader can delve into the following subjects: present cryptographic schemes (symmetric and asymmetric), differences between quantum and classical computing, challenges in quantum computing, quantum algorithms (Shor's and Grover's), public key encryption schemes affected, symmetric schemes affected, the impact on hash functions, and post quantum cryptography. Specifically, the section of Post-Quantum Cryptography deals with different quantum key distribution methods and mathematical-based solutions, such as the BB84 protocol, lattice-based cryptography, multivariate-based cryptography, hash-based signatures and code-based cryptography.
\end{abstract}


\begin{IEEEkeywords}
quantum computers; post-quantum cryptography; Shor's algorithm; Grover's algorithm; asymmetric cryptography; symmetric cryptography
\end{IEEEkeywords}

\section{Introduction}
\label{sec:introduction}
There is no doubt that advancements in technology and particularly electronic communications have become one of the main technological pillars of the modern age. The need for confidentiality, integrity, authenticity, and non-repudiation in data transmission and data storage makes the science of cryptography one of the most important disciplines in information technology. Cryptography, etymologically derived from the Greek words hidden and writing, is the process of securing data in transit or stored by third party adversaries. There are two kinds of cryptosystems; symmetric and asymmetric. 

Quantum computing theory firstly introduced as a concept in 1982 by Richard Feynman, has been researched extensively and is considered the destructor of the present modern asymmetric cryptography. In addition, it is a fact that symmetric cryptography can also be affected by specific quantum algorithms; however, its security can be increased with the use of larger key spaces. Furthermore, algorithms that can break the present asymmetric cryptoschemes whose security is based on the difficulty of factorizing large prime numbers and the discrete logarithm problem have been introduced. It appears that even elliptic curve cryptography which is considered presently the most secure and efficient scheme is weak against quantum computers. Consequently, a need for cryptographic algorithms robust to quantum computations arose.

The rest of the paper deals initially with the analysis of symmetric cryptography, asymmetric cryptography and hash functions. Specifically, an emphasis is given on algorithms that take advantage of the difficulty to factorize large prime numbers, as well as the discrete logarithm problem. We move on by giving an introduction to quantum mechanics and the challenge of building a true quantum computer. Furthermore, we introduce two important quantum algorithms that can have a huge impact in asymmetric cryptography and less in symmetric, namely Shor's algorithm and Grover's algorithm respectively. Finally, post-quantum cryptography is presented. Particularly, an emphasis is given on the analysis of quantum key distribution and some mathematical based solutions such as lattice-based cryptography, multivariate-based cryptography, hash-based signatures, and code-based cryptography.

\section{Present Cryptography}
In this chapter we explain briefly the role of symmetric algorithms, asymmetric algorithms and hash functions in modern cryptography. We analyze the difficulty of factorizing large numbers, as well as the discrete logarithm problem which is the basis of strong asymmetric ciphers.

\subsection{Symmetric Cryptography}
In symmetric cryptography, the sender and the receiver use the same secret key and the same cryptographic algorithm to encrypt and decrypt data. For example, Alice can encrypt a plaintext message using her shared secret key and Bob can decrypt the message using the same cryptographic algorithm Alice used and the same shared secret key. The key needs to be kept secret, meaning that only Alice and Bob should know it; therefore, an efficient way for exchanging secret keys over public networks is demanded. Asymmetric cryptography was introduced to solve the problem of key distribution in symmetric cryptography. Popular symmetric algorithms include the advanced encryption standard (AES) and the data encryption standard (3DES). 

\subsection{Asymmetric Cryptography}
Asymmetric cryptography or public key cryptography (PKC) is a form of encryption where the keys come in pairs. Each party should have its own private and public key. For instance, if Bob wants to encrypt a message, Alice would send her public key to Bob and then Bob can encrypt the message with Alice’s public key. Next, Bob would transmit the encrypted message to Alice who is able to decrypt the message with her private key. Thus, we encrypt the message with a public key and only the person who owns the private key can decrypt the message.

Asymmetric cryptography additionally is used for digital signatures. For example, Alice can sign a document digitally with her private key and Bob can verify the signature with Alice's known public key.
The security of PKC rests on computational problems such as the difficulty of factorizing large prime numbers and the discrete logarithm problem. Such kind of algorithms are called one-way functions because they are easy to compute in one direction but the inversion is difficult \cite{Dusek2006381}. 

\subsubsection{Factorization Problem - RSA Cryptosystem}

One of the most important public-key schemes is RSA invented by Ronald Rivest, Adi Shamir, and Leonard Adleman in 1977. RSA exploits the difficulty of factorizing bi-prime numbers. According to Paar and Pelzl \cite{Paar2010}, RSA and in general asymmetric algorithms are not meant to replace symmetric algorithms because they are computationally costly. RSA is mainly used for secure key exchange between end nodes and often used together with symmetric algorithms such as AES, where the symmetric algorithm does the actual data encryption and decryption. Kirsch \cite{Kirsch2015} stated that RSA is theoretically vulnerable if a fast factorizing algorithm is introduced or huge increase in computation power can exist. The latter can be achieved with the use of quantum mechanics on computers, known as quantum-computers.  

\subsubsection{Discrete Logarithm Problem (DLP)}
Asymmetric cryptographic systems such as Diffie-Hellman (DH) and Elliptic Curve Cryptography (ECC) are based on DLP. The difficulty of breaking these cryptosystems is based on the difficulty in determining the integer $r$ such that $g^{r} = x\mod p$. The integer $r$ is called the discrete logarithm problem of $x$ to the base $g$, and we can write it as $r = \log_{g} x\mod p$. The discrete logarithm problem is a very hard problem to compute if the parameters are large enough. 

Diffie-Hellman is an asymmetric cipher that uses the aforementioned property to transmit keys securely over a public network. Recently, keys larger or equal to $2048$ bits are recommended for secure key exchange. In addition, another family of public key algorithms known as Elliptic Curve Cryptography is extensively used. ECC provides the same level of security as RSA and DLP systems with shorter key operands which makes it convenient to be used by systems of low computational resources. ECC uses a pair $(x,y)$ that fits into the equation $y^2=x^3+ax+b\mod p$ together with an imaginary point $\Theta$ (theta) at infinity, where $a,b \in Z_{p}$ and $4a^3+27b^2 \neq 0\mod p$ \cite{Paar2010}. ECC needs a cyclic Group G and the primitive elements we use, or pair elements, to be of order G. ECC is considered the most secure and efficient asymmetric cryptosystem, but this tends to change with the introduction of quantum computers as it is explained in the next sections.

\section{Quantum Computing vs Classical Computing} \label{Quantum_computing}
In 1982, Richard Feynman came up with the idea of \textit{quantum computer}, a computer that uses the effects of quantum mechanics to its advantage. Quantum mechanics is related to microscopic physical phenomena and their strange behavior. In a traditional computer the fundamental blocks are called bits and can be observed only in two states; 0 and 1. Quantum computers instead use quantum bits also usually referred as \textit{qubits} \cite{Nielsen:2011:QCQ:1972505}. In a sense, qubits are particles that can exist not only in the 0 and 1 state but in both  simultaneously, known as superposition. A particle collapses into one of these states when it is inspected. Quantum computers take advantage of this property mentioned to solve complex problems. An operation on a qubit in superposition acts on both values at the same time. 
Another physical phenomenon used in quantum computing is quantum entanglement. When two qubits are entangled their quantum state can no longer be described independently of each other, but as a single object with four different states. In addition, if one of the two qubits state change the entangled qubit will change too regardless of the distance between them. This leads to true parallel processing power \cite{Quantum_entanglement}.
The combination of the aforementioned phenomena result in exponential increase in the number of values that can be processed in one operation, when the number of entanglement qubits increase. Therefore, a \textit{n-qubit} quantum computer can process $2^n$ operations in parallel.

Two kinds of quantum computers exists; universal and non-universal. The main difference between the two is that universal quantum computers are developed to perform any given task, whereas non-universal quantum computers are developed for a given purpose (e.g., optimization of machine learning algorithms).
Examples are, D-Wave's 2000$+$ qubits non-universal quantum computer \cite{D-wave_interview} and IBM's 17 qubits universal quantum computer with proper error correction. IBM's quantum computer is currently the state of the art of universal quantum computers \cite{soeken2018programming}. Both D-Wave and IBM have quantum computers accessible online for research purposes. Additionally, in October 2017, Intel in collaboration with QuTech announced their 17-qubits universal quantum computer \cite{soeken2018programming}.

Bone and Castro \cite{Bone1997} stated that a quantum computer is completely different in design than a classical computer that uses the traditional transistors and diodes. Researchers have experimented with many different designs such as quantum dots which are basically electrons being in a superposition state, and computing liquids. Besides, they remarked that quantum computers can show their superiority over the classical computers only when used with algorithms that exploit the power of quantum parallelism. For example, a quantum computer would not be any faster than a traditional computer in multiplication. 

\subsection{Challenges in Quantum Computing}
There are many challenges in quantum computing that many researchers are working on.
\begin{itemize} 
\item Quantum algorithms are mainly probabilistic. This means that in one operation a quantum computer returns many solutions where only one is the correct. This trial and error for measuring and verifying the correct answer weakens the advantage of quantum computing speed \cite{Kirsch2015}.

\item Qubits are susceptible to errors. They can be affected by heat, noise in the environment, as well as stray electromagnetic couplings. Classical computers are susceptible to bit-flips (a zero can become one and vise versa). Qubits suffer from bit-flips as well as phase errors. Direct inspection for errors should be avoided as it will cause the value to collapse, leaving its superposition state. 

\item Another challenge is the difficulty of coherence. Qubits can retain their quantum state for a short period of time. Researchers at the University of New South Wales in Australia have created two different types of qubits (Phosphorous atom and an Artificial atom) and by putting them into a tiny silicon (\textit{silicon 28}) they were able to elliminate the magnetic noise that makes them prone to errors. Additionally, they stated that the Phosphorous atom has 99.99\% accuracy which accounts for 1 error every 10,000 quantum operations \cite{Muhonen2014}. Their qubits can remain in superposition for a total of 35 seconds which is considered a world record \cite{D-Wave2016}. Moreover, to achieve long coherence qubits need not only to be isolated from the external world but to be kept in temperatures reaching the absolute zero. However, this isolation makes it difficult to control them without contributing additional noise \cite{Kirsch2015}.
\end{itemize}
IBM in 2017, introduced the definition of \textit{Quantum Volume}. Quantum volume is a metric to measure how powerful a quantum computer is based on how many qubits it has, how good is the error correction on these qubits, and the number of operations that can be done in parallel. Increase in the number of qubit does not improve a quantum computer if the error rate is high. However, improving the error rate would result in a more powerful quantum computer \cite{Quantum_volume}.

\section{Cryptosystems Vulnerable to Quantum Algorithms}

This section discusses the impact of quantum algorithms on present cryptography and gives an introduction to  Shor's algorithm and Grover's algorithm. 
Note that Shor's algorithm explained in the following subsection makes the algorithms that rely on the difficulty of factorizing or computing discrete logarithms vulnerable. 

Cryptography plays an important role in every electronic communication system today. For example the security of emails, passwords, financial transactions, or even electronic voting systems require the same security objectives such as confidentiality and integrity \cite{Campagna2015}. Cryptography makes sure that only parties that have exchanged keys can read the encrypted message (also called authentic parties). Quantum computers threaten the main goal of every secure and authentic communication because they are able to do computations that classical (conventional) computers cannot. Consequently, quantum computers can break the cryptographic keys quickly by calculating or searching exhaustively all secret keys, allowing an eavesdropper to intercept the communication channel between authentic parties (sender/receiver). This task is considered to be computational infeasible by a conventional computer \cite{Buchanan2016}. 

\begin{table*}[h]
\centering
\caption{Impact analysis of quantum computing on encryption schemes (adapted from \cite{Chen2016})}
\label{table1}
\begin{adjustbox}{width=1\textwidth}
\begin{tabular}{|
>{\columncolor[HTML]{EFEFEF}}p{5cm} |
>{\columncolor[HTML]{EFEFEF}}l |
>{\columncolor[HTML]{EFEFEF}}p{3cm} |
>{\columncolor[HTML]{EFEFEF}}p{3cm} |}
\Xhline{1\arrayrulewidth}
\textbf{Cryptographic Algorithm}          & \textbf{Type} & \textbf{Purpose}              & \textbf{Impact From Quantum Computer} \\ \Xhline{1\arrayrulewidth}
AES-256                                   & Symmetric key & Encryption                    & Secure               \\ \Xhline{1\arrayrulewidth}
SHA-256, SHA-3                            & --            & Hash functions                & Secure                  \\ \Xhline{1\arrayrulewidth}
RSA                                       & Public key    & Signatures, key establishment & No longer secure                      \\ \Xhline{1\arrayrulewidth}
ECDSA, ECDH (Elliptic Curve Cryptography) & Public key    & Signatures, key exchange      & No longer secure                      \\ \Xhline{1\arrayrulewidth}
DSA (Finite Field Cryptography)           & Public key    & Signatures, key exchange      & No longer secure                      \\ \Xhline{1\arrayrulewidth}
\end{tabular}
\end{adjustbox}
\end{table*}

According to NIST, quantum computers will bring the end of the current public key encryption schemes \cite{Chen2016}. Table \ref{table1} adapted from NIST shows the impact of quantum computing on present cryptographic schemes.

\subsection{Shor's Algorithm in Asymmetric Cryptography}
In 1994, the mathematician Peter Shor in his paper “Algorithms for Quantum Computation: Discrete Logarithms and Factoring” \cite{Shor1994}, proved that factorizing large integers would change fundamentally with a quantum computer.

Shor's algorithm can make modern asymmetric cryptography collapse since is it based on large prime integer factorization or the discrete logarithm problem. To understand how Shor's algorithm factorizes large prime numbers we use the following example. We want to find the prime factors of number 15. To do so, we need a 4-qubit register. We can visualize a 4-qubit register as a normal 4-bit register of a traditional computer. Number 15 in binary is 1111, so a 4-qubit register is enough to accommodate (calculate) the prime factorization of this number. According to Bone and Castro \cite{Bone1997}, a calculation performed on the register can be thought as computations done in parallel for every possible value that the register can take (0-15). This is also the only step needed to be performed on a quantum computer. 

The algorithm does the following: 

\begin{itemize}
\item n = 15, is the number we want to factorize
\item x = random number such as $1<x<n-1$
\item x is raised to the power contained in the register (every possible state) and then divided by \textit{n}

The remainder from this operation is stored in a second 4-qubit register. The second register now contains the superposition results. Let's assume that $x=2$ which is larger than 1 and smaller than 14. 

\item If we raise $x$ to the powers of the 4-qubit register which is a maximum of 15 and divide by 15, the remainders are shown in Table \ref{tab:registers}.

\begin{table*}[h]
\centering
 \caption{4-qubit registers with remainders}
 \label{tab:registers}
\resizebox{1\textwidth}{!}{%
\begin{tabular}{
>{\columncolor[HTML]{EFEFEF}}l 
>{\columncolor[HTML]{EFEFEF}}l 
>{\columncolor[HTML]{EFEFEF}}l 
>{\columncolor[HTML]{EFEFEF}}l 
>{\columncolor[HTML]{EFEFEF}}l 
>{\columncolor[HTML]{EFEFEF}}l 
>{\columncolor[HTML]{EFEFEF}}l 
>{\columncolor[HTML]{EFEFEF}}l 
>{\columncolor[HTML]{EFEFEF}}l 
>{\columncolor[HTML]{EFEFEF}}l 
>{\columncolor[HTML]{EFEFEF}}l 
>{\columncolor[HTML]{EFEFEF}}l 
>{\columncolor[HTML]{EFEFEF}}l 
>{\columncolor[HTML]{EFEFEF}}l 
>{\columncolor[HTML]{EFEFEF}}l 
>{\columncolor[HTML]{EFEFEF}}l 
>{\columncolor[HTML]{EFEFEF}}l }
\textbf{Register 1:} & 0                                              & 1                                              & 2                                              & 3                                              & 4                                              & 5                                              & 6                                              & 7                                              & 8                                              & 9                                              & 10                                             & 11                                             & 12                                             & 13                                             & 14                                             & 15                                             \\ \hline
\textbf{Register 2:} & \multicolumn{1}{c}{\cellcolor[HTML]{EFEFEF}1} & \multicolumn{1}{c}{\cellcolor[HTML]{EFEFEF}2} & \multicolumn{1}{c}{\cellcolor[HTML]{EFEFEF}4} & \multicolumn{1}{c}{\cellcolor[HTML]{EFEFEF}8} & \multicolumn{1}{c}{\cellcolor[HTML]{EFEFEF}1} & \multicolumn{1}{c}{\cellcolor[HTML]{EFEFEF}2} & \multicolumn{1}{c}{\cellcolor[HTML]{EFEFEF}4} & \multicolumn{1}{c}{\cellcolor[HTML]{EFEFEF}8} & \multicolumn{1}{c}{\cellcolor[HTML]{EFEFEF}1} & \multicolumn{1}{c}{\cellcolor[HTML]{EFEFEF}2} & \multicolumn{1}{c}{\cellcolor[HTML]{EFEFEF}4} & \multicolumn{1}{c}{\cellcolor[HTML]{EFEFEF}8} & \multicolumn{1}{c}{\cellcolor[HTML]{EFEFEF}1} & \multicolumn{1}{c}{\cellcolor[HTML]{EFEFEF}2} & \multicolumn{1}{c}{\cellcolor[HTML]{EFEFEF}4} & \multicolumn{1}{c}{\cellcolor[HTML]{EFEFEF}8} \\ 
\end{tabular}}
\end{table*}

What we observe in the results is a repeating sequence of 4 numbers (1,2,4,8). We can confidently say then that $f=4$ which is the sequence when $x=2$ and $n=15$. The value $f$ can be used to calculate a possible factor with the following equation:

\textbf{Possible factor:} $P = x^{f/2}-1$
\end{itemize}

In case we get a result which is not a prime number we repeat the calculation with different $f$ values.

Shor’s algorithm can be used additionally for computing discrete logarithm problems. Vazirani \cite{Vazirani1997} explored in detail the methodology of Shor’s algorithm and showed that by starting from a random superposition state of two integers, and by performing a series of Fourier transformations, a new superposition can be set-up to give us with high probability two integers that satisfy an equation. By using this equation we can calculate the value $r$ which is the unknown "exponent" in the DLP.

\subsection{Grover's algorithm in Symmetric Cryptography}
Lov Grover created an algorithm that uses quantum computers to search unsorted databases \cite{Grover1996}. The algorithm can find a specific entry in an unsorted database of $N$ entries in $\sqrt{N}$ searches. In comparison, a conventional computer would need $N/2$ searches to find the same entry.
Bone and Castro \cite{Bone1997} remarked the impact of a possible application of Grover’s algorithm to crack Data Encryption Standard (DES), which relies its security on a 56-bit key. The authors remarked that the algorithm needs only 185 searches to find the key. 

Currently, to prevent password cracking we increase the number of key bits (larger key space); as a result, the number of searches needed to crack a password increases exponentially. 
Buchmann et al. \cite{Buchmann2010} stated that Grover’s algorithm have some applications to symmetric cryptosystems but it is not as fast as Shor’s algorithm. 

\subsection{Asymmetric Encryption Schemes Affected}
All public key algorithms used today are based on two mathematical problems, the aforementioned factorization of large numbers (e.g., RSA) and the calculation of discrete logarithms (e.g., DSA signatures and ElGamal encryption). Both have similar mathematical structure and can be broken with Shor's algorithm rapidly. Recent algorithms based on elliptic curves (such as ECDSA) use a modification of the discrete logarithm problem that makes them equally weak against quantum computers. Kirsch and Chow \cite{Kirsch2015} mentioned that a modified Shor's algorithm can be used to decrypt data encrypted with ECC. In addition, they emphasized that the relatively small key space of ECC compared to RSA makes it easier to be broken by quantum computers. Furthermore, Proos and Zalka \cite{Proos2003} explained that 160-bit elliptic curves could be broken by a 1000-qubit quantum computer, while factorizing 1024-bit RSA would require a 2000-qubit quantum computer. The number of qubits needed to break a cryptosystem is relative to the algorithm proposed. In addition, they show in some detail how to use Shor’s algorithm to break ECC over GF(p).

On the other hand, Grover’s algorithm is a threat only to some symmetric cryptographic schemes. NIST \cite{Chen2016} points out that if the key sizes are sufficient, symmetric cryptographic schemes (specifically the Advanced Encryption Standard-AES) are resistant to quantum computers. Another aspect to be taken into consideration is the robustness of algorithms against quantum computing attacks also known as quantum cryptanalysis. 

In table \ref{table2}, a comparison of classical and quantum security levels for the most used cryptographic schemes is presented.

\subsection{Symmetric Encryption Schemes Affected}
For symmetric cryptography quantum computing is considered a minor threat. The only known threat is Grover's algorithm that offers a square root speed-up over classical brute force algorithms. For example, for a n-bit cipher the quantum computer operates on ($\sqrt{2^{n}} = 2^{n/2}$). In practice, this means that a symmetric cipher with a key length of 128-bit (e.g., AES-128) would provide a security level of 64-bit. 
We recall here that security level of 80-bit is considered secure. The Advanced Encryption Standard (AES) is considered to be one of the cryptographic primitives that is resilient in quantum computations, but only when is used with key sizes of 192 or 256 bits. Another indicator of the security of AES in the post-quantum era is that NSA (The National Security Agency) allows AES cipher to secure (protect) classified information for security levels, SECRET and TOP SECRET, but only with key sizes of 192 and 256 bits \cite{NationalSecurityAgency2003}.

\begin{table}[h]
\centering
\caption{Comparison of classical and quantum security levels for the most used cryptographic schemes}
\label{table2}
\resizebox{0.47\textwidth}{!}{%
\begin{tabular}{|
>{\columncolor[HTML]{ffffff}}l |
>{\columncolor[HTML]{ffffff}}l |
>{\columncolor[HTML]{EFEFEF}}c |
>{\columncolor[HTML]{EFEFEF}}c |}
\hline
\cellcolor[HTML]{ffffff}                                         & \cellcolor[HTML]{ffffff}                                    & \multicolumn{2}{l|}{\cellcolor[HTML]{EFEFEF}\textbf{Effective Key Strength/Security Level (in bits)}}                                               \\ \hhline{~~--}
\multirow{-2}{*}{\cellcolor[HTML]{ffffff}\textbf{Crypto Scheme}} & \multirow{-2}{*}{\cellcolor[HTML]{ffffff}\textbf{Key Size}} & \multicolumn{1}{l|}{\cellcolor[HTML]{EFEFEF}\textbf{Classical Computing}} & \multicolumn{1}{l|}{\cellcolor[HTML]{EFEFEF}\textbf{Quantum Computing}} \\ \hline
RSA-1024                                                         & 1024                                                        & 80                                                                        & 0                                                                       \\ \hline
RSA-2048                                                         & 2048                                                        & 112                                                                       & 0                                                                       \\ \hline
ECC-256                                                          & 256                                                         & 128                                                                       & 0                                                                       \\ \hline
ECC-384                                                          & 384                                                         & 256                                                                       & 0                                                                       \\ \hline
\textbf{AES-128}                                                 & \textbf{128}                                                & \textbf{128}                                                              & \textbf{64}                                                             \\ \hline
\textbf{AES-256}                                                 & \textbf{256}                                                & \textbf{256}                                                              & \textbf{128}                                                            \\ \hline
\end{tabular}}
\end{table}

\subsection{Hash Functions}
The family of hash functions suffer from a similar problem as symmetric ciphers since their security depends on a fixed output length. Grover's algorithm can be utilized to find a collision in a hash function in square root steps of its original length (it is like searching an unsorted database). In addition, it has been proved that it is possible to combine Grover's algorithm with the birthday paradox. Brassard et al. \cite{Brassard1998} described a quantum birthday attack. By creating a table of size $\sqrt[3]{N}$ and utilizing Grover’s algorithm to find a collision an attack is said to work effectively. This means that to provide a $b-bit$ security level against Grover's quantum algorithm a hash function must provide at least a $3b-bit$ output. As a result, many of the present hash algorithms are disqualified for use in the quantum era. However, both SHA-2 and SHA-3 with longer outputs, remain quantum resistant.

\section{Post-Quantum Cryptography}
The goal of post-quantum cryptography (also known as quantum-resistant cryptography) is to develop cryptographic systems that are secure against both quantum and conventional computers and can interoperate with existing communication protocols and networks \cite{Chen2016}. Many post-quantum public key candidates are actively investigated the last years. In 2016, NIST announced a call for proposals of algorithms that are believed to be quantum resilient with a deadline in November 2017. In January 2018, NIST published the results of the first round. In total 82 algorithms were proposed from which 59 are encryption or key exchange schemes and 23 are signature schemes. After 3 to 5 years of analysis NIST will report the findings and prepare a draft of standards \cite{Asiacrypt_2017_Moody}. Furthermore, the National Security Agency (NSA) has already announced plans to migrate their cryptographic standards to post-quantum cryptography \cite{ArsTechnica2015}.

The cryptographic algorithms presented in this section do not rely on the hidden subgroup problem (HSP) such as factorizing integers or computing discrete logarithms, but different complex mathematical problems.

\subsection{Quantum Key Distribution}
Quantum Key Distribution (QKD) addresses the challenge of securely exchanging a cryptographic key between two parties over an insecure channel. QKD relies on the fundamental characteristics of quantum mechanics which are invulnerable to increasing computational power, and may be performed by using the quantum properties of light, lasers, fibre-optics as well as free space transmission technology. QKD was first introduced in 1984 when Charles Bennett and Gilles Brassard developed their BB84 protocol \cite{BennettCharlesandBrassard1984,Bennett1992}. Research has led to the development of many new QKD protocols exploiting mainly two different properties that are described right below.

Prepare-and-measure (P\&M) protocols use the Heisenberg Uncertainty principle \cite{panarella1987heisenberg} stating that the measuring act of a quantum state changes that state in some way. This makes it difficult for an attacker to eavesdrop on a communication channel without leaving any trace. In case of eavesdropping the legitimate exchange parties are able to discard the corrupted information as well as to calculate the amount of information that has been intercepted \cite{singh2012quantum}. This property was exploited in BB84.

Entanglement based (EB) protocols use pairs of entangled objects which are shared between two parties. As explained in \ref{Quantum_computing}, entanglement is a quantum physical phenomenon which links two or more objects together in such a way that afterwards they have to be considered as one object. Additionally, measuring one of the objects would affect the other as well. In practice when an entangled pair of objects is shared between two legitimate exchange parties anyone intercepting either object would alter the overall system. This would reveal the presence of an attacker along with the amount of information that the attacker retrieved. This property was exploited in E91 \cite{ekert1991quantum} protocol. 

Both of the above-mentioned approaches are additionally divided into three families; discrete variable coding, continuous variable coding and distributed phase reference coding. The main difference between these families is the type of detecting system used. Both discrete variable coding and distributed phase reference coding use photon counting and post-select the events in which a detection has effectively taken place \cite{scarani2009security}. Continuous variable coding uses homodyne detection \cite{scarani2009security} which is a comparison of modulation of a single frequency of an oscillating signal with a standard oscillation.

A concise list of QKD protocols for the aforementioned families is presented below.

Discrete variable coding protocols:
\begin{itemize}
    \item BB84 \cite{BennettCharlesandBrassard1984,Bennett1992} - the first QKD protocol that uses four non-orthogonal polarized single photon states or low-intensity light pulses. A detailed description of this protocol is given below.
    \item BBM \cite{bennett1992quantum_Bassard} - is an entanglement based version of BB84.
    \item E91 \cite{ekert1991quantum} - is based on the \textit{gedanken experiment} \cite{bohm1951quantum} and the generalized Bell's theorem \cite{clauser1969proposed}. In addition, it can be considered an extension of Bennett and Brassard's (authors of BB84) original idea. 
    \item SARG04  \cite{scarani2004quantum,acin2004coherent} - is similar to BB84 but instead of using the state to code the bits, the bases are used. SARG04 is more robust than BB84 against the photon number splitting (PNS) attack.

\item Six state protocol \cite{bennett1984g,bruss1998optimal,bechmann1999incoherent} - is a version of BB84 that uses a six-state polarization scheme on three orthogonal bases. 
 \item Six state version of the SARG04 coding \cite{tamaki2006unconditionally}.
\item Singapore protocol \cite{englert2004efficient} -  is a tomographic protocol that is more efficient than the Six state protocol. 
\item B92 protocol \cite{bennett1992quantum} - two non-orthogonal quantum states using low-intensity coherent light pulses.\\
    \end{itemize}
    
Continuous variable coding protocols:
\begin{itemize}
\item Gaussian protocols
\begin{itemize}
    \item Continuous variable version of BB84 \cite{cerf2001quantum}
    \item Continuous variable using coherent states \cite{grosshans2002continuous}
    \item Coherent state QKD protocol \cite{weedbrook2004quantum} - based on simultaneous quadrature measurements.
    \item Coherent state QKD protocol \cite{lodewyck2007quantum} - based on the generation and transmission of random distributions of coherent or squeezed states.
\end{itemize}
\item Discrete-modulation protocols
\begin{itemize}
\item First continuous variable protocol based on coherent states instead of squeezed states \cite{silberhorn2002continuous}.
\end{itemize}
\end{itemize}
Distributed phase reference coding protocols:
\begin{itemize}
\item Differential Phase Shift (DPS) Quantum Key Distribution (QKD) protocol \cite{inoue2002differential,inoue2003differential} - uses a single photon in superposition state of three basis kets, where the phase difference between two sequential pulses carries bit information.
\item Coherent One Way (COW) protocol \cite{gisin2004towards,stucki2005fast} - 
 the key is obtained by a time-of-arrival measurement on the data line (raw key). Additionally, an interferometer is built on a monitoring line, allowing to monitor the presence of an intruder. A prototype was presented in 2008 \cite{stucki2009continuous}.
\end{itemize}
Discrete variable coding protocols are the most widely implemented, whereas the continuous variable and distributed phase reference coding protocols are mainly concerned with overcoming practical limitations of experiments.

\subsubsection{BB84 protocol} \label{BB84}
BB84 is the first quantum cryptographic protocol (QKD scheme) which is still in use today. According to Mayers \cite{Mayers2001} BB84 is \textit{provable secure}, explaining that a secure key sequence can be generated whenever the channel bit error rate is less than about 7\% \cite{Branciard}. BB84 exploits the polarization of light for creating random sequence of qubits (key) that are transmitted through a quantum channel.

\begin{figure*}[h]
	\centering
		\includegraphics[width=0.9\textwidth]{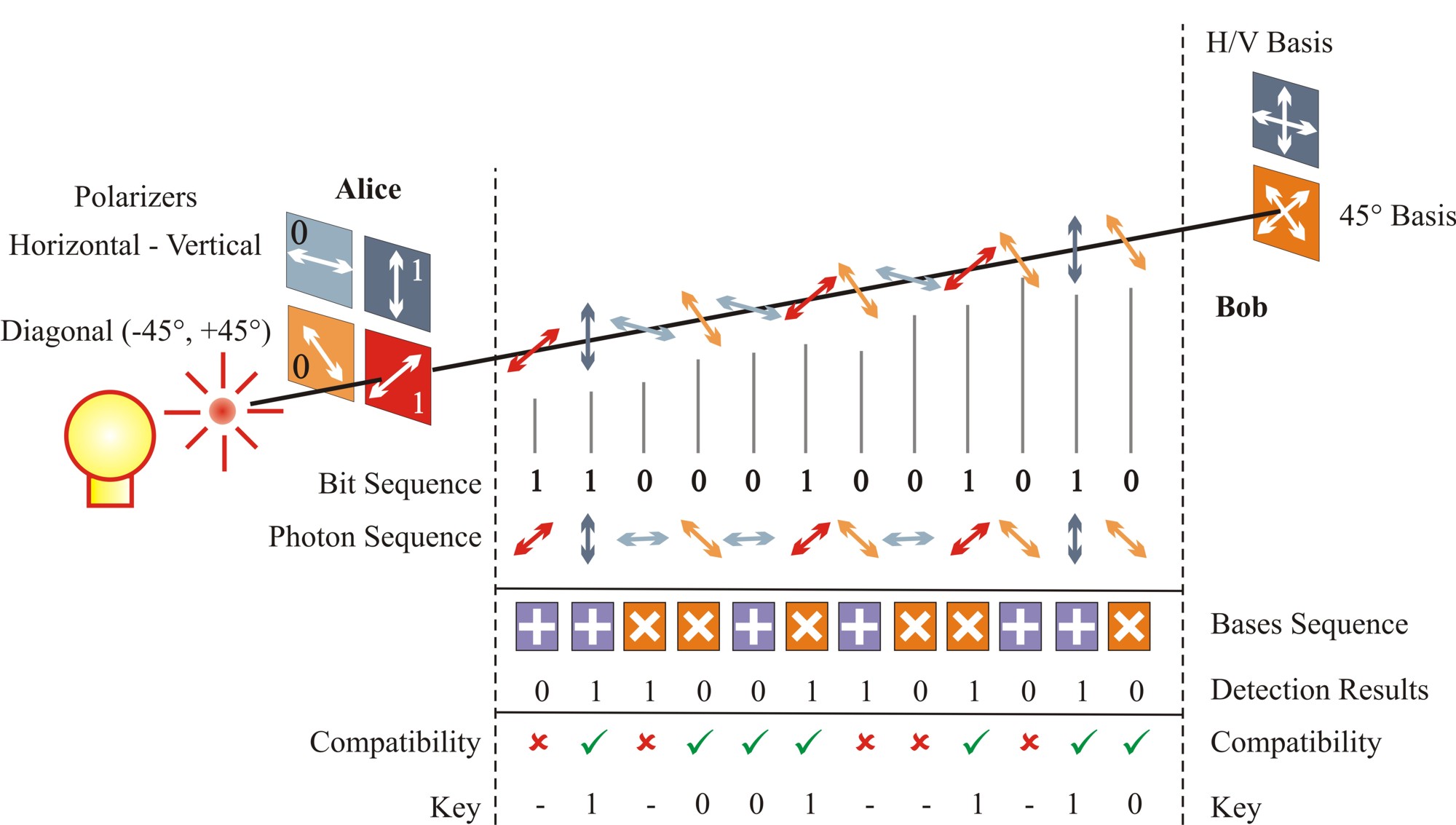}
	\caption{Key exchange in the BB84 protocol implemented with polarization of photons (adapted from \cite{Makarov2007}).}
 	\label{fig:bb84}
 \end{figure*}

BB84 uses two different bases, base 1 is polarized $0^{o}$ (horizontal) or $90^{o}$ (vertical) with $0^{o}$ equal to 0 and $90^{o}$ equal to 1. Base 2 is polarized $45^{o}$ or $135^{o}$ with $45^{o}$ equal to 1 and $135^{o}$ equal to 0. Alice begins by sending a photon in one of the two bases having a value of 0 or 1. Both the base and the value should be chosen randomly. Next, Bob selects the base 1 or 2 and measures a value without knowing which base Alice has used. The key exchange process continues until they have generated enough bits. Furthermore, Bob tells Alice the sequence of the bases he used but not the values he measured and Alice informs Bob whether the chosen bases were right or wrong. If the base is right, Alice and Bob have equal bits, whereas if it is wrong the bits are discarded. In addition, any bits that did not make it to the destination are discarded by Alice. Now Alice can use the key that they just exchanged to encode the message and send it to Bob. BB84 is illustrated visually in Figure \ref{fig:bb84}.

Worthy to mentioning is that this method of communication was broken by Lydersen et al. in 2010 \cite{QKD_hacked}. Their experiment proved that although BB84 is \textit{provable secure} the actual hardware implemented is not. 
The authors managed to inspect the secret key without the receiver noticing it by blinding the APD-based detector (avalanche photodiode).

Yuan et al. \cite{Avoid_QKD_hacking} proposed improvements to mitigate blinding attacks, such as monitoring the photocurrent for anomalously high values. Lydersen et al. \cite{Reply_to_Avoid_QKD_hacking} after taking into consideration the improvements of Yuan et al. \cite{Avoid_QKD_hacking} succeeded again to reveal the secret key without leaving any traces. 

\subsubsection{Photon Number Splitting Attack}
The crucial issue in quantum key distribution is its security. In addition to noise in the quantum channel, the equipment is impractical to produce and detect single photons. Therefore, in practice, laser pulses are used. Producing multiple photons opens up a new attack known as Photon Number Splitting (PNS) attack. In PNS attack, an attacker (Eve) deterministically splits a photon off of the signal and stores it in a quantum memory which does not modify the polarisation of the photons. The remaining photons are allowed to pass and are transmitted to the receiver (Bob). Next, Bob measures the photons and the sender (Alice) has to reveal the encoding bases. Eve will then be able to measure all captured photons on a correct bases. Consequently, Eve will obtain information about the secret key from all signals containing more than one photon without being noticed \cite{brassard2000security}.

Different solutions have been proposed for mitigating PNS attacks. The most promising solution developed by Lo et al. \cite{lo2005decoy} uses decoy states to detect PNS attacks. This is achieved by sending  randomly laser pulses with a lower average photon number. Thereafter, Eve cannot distinguish between decoyed signals and non-decoyed signals. This method works for both single and multi-photon pulses \cite{haitjema2007survey}.  

\subsection{Mathematically-based Solutions}
There are many alternative mathematical problems to those used in RSA, DH and ECDSA that have already been implemented as public key cryptographic schemes, and for which the Hidden Subgroup Problem (HSP) \cite{Lomonaco2002} does not apply; therefore, they appear to be quantum resistant. 

The most researched mathematical-based implementations are the following:

\begin{itemize}
\item Lattice-based cryptography \cite{Micciancio2009} 
\item Multivariate-based cryptography \cite{Ding2009} 
\item Hash-based signatures \cite{Dods2005}
\item Code-based cryptography \cite{Overbeck2009} 
\end{itemize}

The existing alternatives and new schemes emerging from these areas of mathematics do not all necessarily satisfy the characteristics of an ideal scheme. In the following subsections we are going to give an overview of these cryptographic schemes.

\subsubsection{Lattice-based Cryptography}
This is a form of public-key cryptography that avoids the weaknesses of RSA. Rather than multiplying primes, lattice-based encryption schemes involve multiplying matrices. Furthermore, lattice-based cryptographic constructions are based on the presumed hardness of lattice problems, the most basic of which is the shortest vector problem (SVP) \cite{Micciancio2009}. Here, we are given as input a lattice represented by an arbitrary basis and our goal is to output the shortest non-zero vector in it.

The Ajtai-Dwork (AD) \cite{Ajtai1997}, Goldreich-Goldwasser-Halevi (GGH) \cite{Goldreich1997} and NTRU \cite{Hoffstein1998} encryption schemes that are explained below are lattice-based cryptosystems.

In 1997, Ajtai and Dwork\cite{Ajtai1997} found the first connection between the worst and the average case complexity of the Shortest Vector Problem (SVP). They claimed that their cryptosystem is \textit{provably secure}, but in 1998, Nguyen and Ster \cite{Nguyen1998} refuted it. Furthermore, the AD public key is big and it causes message expansion making it an unrealistic public key candidate in post-quantum era. 

The Goldreich-Goldwasser-Halevi (GGH) was published in 1997. GGH makes use of the Closest Vector Problem (CVP) which is known to be NP-hard. Despite the fact that GGH is more efficient than Ajtai-Dwork (AD), in 1999, Nguyen\cite{Nguyen1999} proved that GGH has a major flaw; partial information on plaintexts can be recovered by solving CVP instances.

NTRU was published in 1996 by Hoffstein et al. \cite{Hoffstein1998}. It is used for both encryption (\textit{NTRUEncrypt}) and digital signature (\textit{NTRUSign}) schemes. NTRU relies on the difficulty of factorizing certain polynomials making it resistant against Shor's algorithm. To provide 128-bit post-quantum security level NTRU demands 12881-bit keys \cite{Hoffstein}. As of today there is not any known attack for NTRU.

In 2013, Damien Stehle and Ron Steinfeld developed a \textit{provably secure} version of NTRU (SS-NTRU) \cite{cryptoeprint2013}.

In May 2016, Bernstein et al. \cite{bernstein2016ntru} released a new version of NTRU called "NTRU Prime". NTRU Prime countermeasures the weaknesses of several lattice based cryptosystems, including NTRU, by using different more secure ring structures.

In conclusion, among all the lattice-based candidates mentioned above NTRU is the most efficient and secure algorithm making it a promising candidate for the post-quantum era.

\subsubsection{Multivariate-based Cryptography}
The security of this public key scheme relies on the difficulty of solving systems of multivariate polynomials over finite fields. Research has shown that development of an encryption algorithm based on multivariate equations is difficult \cite{Buchanan2016}. Multivariate cryptosystems can be used both for encryption and digital signatures. Tao et al. \cite{tao2013simple} explained  that there have been several attempts to build asymmetric pubic key encryption schemes based on multivariate polynomials; however, most of them are insecure because of the fact that certain quadratic forms associated with their central maps have low rank. The authors \cite{tao2013simple} proposed a new efficient multivariate scheme, namely Simple Matrix (ABC), based on matrix multiplication that overcomes the aforementioned weakness. In addition, multivariate cryptosystems can be used for digitals signatures. The most promising signature schemes include Unbalanced Oil and Vinegar (multivariate quadratic equations), and Rainbow. UOV has a large ratio between the number of variables and equations (3:1) making the signatures three times longer than the hash values. In addition, the public key sizes are large. On the other hand, Rainbow is more efficient by using smaller ratios which result in smaller digital signatures and key sizes \cite{Campagna2015}.

\subsubsection{Hash-based Signatures}

In this subsection, we introduce the Lamport signature scheme invented in 1979 by Leslie Lamport. Buchmann et al. \cite{Buchmann2010} introduced concisely the scheme. A parameter $b$ defines the desired security level of our system. For 128-bit $b$ security level we need a secure hash function that takes arbitrary length input and produces 256-bit length output; thus, SHA-256 is considered an optimal solution that can be fitted with our message $m$.

\textbf{Private key:} A random number generator is used to produce 256 pairs of random numbers. Each number is 256 bits. In total our generated numbers are $2\times256\times256 = 16$ KB. Therefore, we can precisely say that the private key consists of $8b^2$ bits.

\textbf{Public key:} All generated numbers (private key) are hashed independently creating 512 different hashes (256 pairs) of 256-bit length each. Therefore, we can precisely say that the public key consists of $8b^2$ bits.

The next step is to sign the message. We have a hashed message \textit{m} and then for each bit (depending on its value 0 or 1) of the message digest we choose one number from each pair that comprise the private key. As a result, we have a sequence of 256 numbers (relative to the bit sequence of the hashed message \textit{m}). The sequence of numbers is the digital signature published along with the plaintext message. It is worth noting that the private key should never be used again and the remaining 256 numbers from the pairs should be destroyed \textit{(Lamport one-time signature)}.

The verification process is straightforward. The recipient calculates the hash of the message and then, for each bit of the hashed message we choose the corresponding hash from the public key (512 in number). In addition, the recipient hashes each number of the sender's private key which should correspond to the same sequence of hashed values with the recipients correctly chosen public key values.
The security of this system derives by the decision of using the private key only once. 
Consequently, an adversary can only retrieve 50 percent of the private key which makes it impossible to forge a new valid signature.

Buchmann et al. \cite{Buchmann2010} explained that in case we want to sign more than one messages, chaining can be introduced. The signer includes in the signed message a newly generated public key that is used to verify the next message received.

Witernitz described a one time signature (WOTS) which is more efficient than Lamport's. Specifically, the signature size and the keys are smaller \cite{Improved_L-OTS_by_Merkle}. However, OTSs are not suitable for large-scale use because they can be used only once.

Merkle introduced a new approach that combines Witernitz's OTS with binary trees (Merkle Signature Scheme). A binary tree is made of nodes. In our case each node represents the hash value of the concatenation of the child nodes. Each of the leaf nodes (lowest nodes in the tree hierarchy) contains a Witernitz's OTS which is used for signing. The first node in the hierarchy of the tree known as root node is the actual public key that can verify the OTSs contained in the leaf nodes \cite{Improved_L-OTS_by_Merkle}. 

In 2013, A. Hulsing improved the WOTS algorithm by making it more efficient without affecting its security level even when hash functions without collision resistance are used \cite{Hulsing2013_W_OTS+}. 

Currently two hash-based signature schemes are under evaluation for standardization. Specifically, the eXtended Merkle Signature Scheme (XMSS) \cite{buchmann2011xmss} which is a stateful signature scheme, and Stateless Practical Hash-based Incredibly Nice Collision-resilient Signatures (SPHINCS) \cite{SPHINCS} which is as the name indicates a stateless signature scheme. 

\subsubsection{Code-based Cryptography}

Code-based cryptography refers to cryptosystems that make use of error correcting codes. The algorithms are based on the difficulty of decoding linear codes and are considered robust to quantum attacks when the key sizes are increased by the factor of 4. Furthermore, Buchmann et al. \cite{Buchmann2010} state that the best way to solve the decoding problem is to transform it to a Low-Weight-Code-World Problem (LWCWP) but solving a LWCWP in large dimensions is considered infeasible. It would be easier to comprehend the process of this scheme by using Buchmann's \cite{Buchmann2010} concise explanation of McEliece's original code-based public-key encryption system. We define b as the security of our system and it is a power of 2. $n = 4b \lg b,\quad d = \lg n,\quad\textrm{and}\quad t = 0.5n/d$.

For example, if $b=128$ then $n=512 \log_2 (128)$ which is equal to 3584. $d = 12$ and $t = 149$.
The receiver's public key in this system is $dtn$ matrix K with coefficients $F_2$. Messages to be encrypted should have exactly \textit{t} bits set to 1 and for the encryption the message \textit{m} is multiplied by K. The receiver generates a public key with a hidden Goppa code structure (error-correction code) that allows to decode the message with Patterson’s algorithm, or even by faster algorithms. The code’s generator matrix K is perturbated by two invertible matrices which are used to decrypt the ciphertext to obtain the message \textit{m}.

As for any other class of cryptosystems, the practice of code-based cryptography is a trade-off between efficiency and security. McEliece's cryptosystem encryption and decryption process are fast with very low complexity, but it makes use of large public keys (100 kilobytes to several megabytes).

\section{Conclusion}
In today's world, where information play a particularly important role, the transmission  and the storage of data must be maximally secure. Quantum computers pose a significant risk to both conventional public key algorithms (such as RSA, ElGamal, ECC and DSA) and symmetric key algorithms (3DES, AES). Year by year it seems that we are getting closer to create a fully operational universal quantum computer that can utilize strong quantum algorithms such as Shor's algorithm and Grover's algorithm. The consequence of this technological advancement is the absolute collapse of the present public key algorithms that are considered secure, such as RSA and Elliptic Curve Cryptosystems. The answer on that threat is the introduction of cryptographic schemes resistant to quantum computing, such as quantum key distribution methods like the BB84 protocol, and mathematical-based solutions like lattice-based cryptography, hash-based signatures, and code-based cryptography.

\section*{Acknowledgment}

This research is supported by the Research Council of Norway under the Grant No.: IKTPLUSS 247648 and 248030/O70 for Oslo Analytics and SWAN projects, respectively. This research is also part of the SecurityLab of the University of Oslo.

\bibliographystyle{IEEEtran}
\bibliography{bibliography}

\end{document}